\documentstyle[aps,multicol,psfig]{revtex}

\newcommand{\be}{\begin{equation}}
\newcommand{\ee}{\end{equation}}
\newcommand{\bea}{\begin{eqnarray}}
\newcommand{\eea}{\end{eqnarray}}
\newcommand \ga{\raisebox{-.5ex}{$\stackrel{>}{\sim}$}}
\newcommand \la{\raisebox{-.5ex}{$\stackrel{<}{\sim}$}}
\newcommand{\rx}{{\rm x}}
\newcommand{\ry}{{\rm y}}
\newcommand{\rz}{{\rm z}}
\begin{document}
\draft

\title{Elliptic Flow and HBT in\\ 
non-central Nuclear Collisions}

\author{Henning Heiselberg}

\address{NORDITA, Blegdamsvej 17, DK-2100 Copenhagen \O, Denmark}

\author{Anne-Marie Levy}

\address{Niels Bohr Institute, Blegdamsvej 17, DK-2100 Copenhagen \O, Denmark}

\maketitle

\begin{abstract}
Elliptic flow and HBT are studied for non-central relativistic nuclear
collisions. Azimuthal asymmetries show up in both elliptic flow and
HBT radii and are calculated in both collisionless and hydrodynamic
limits relevant for peripheral and central collisions respectively.
Determining the reaction plane and measuring the HBT radii as function
of the angle between the reaction plane and the particle momenta can
determine the physical quantities as source sizes, deformations,
emission times, duration of emission and opacities. Comparison to SPS
data and predictions for RHIC and LHC energies are given.  The
centrality dependence with and without a
phase transition to a quark-gluon plasma is discussed - in particular,
how the physical quantities are expected to display a
qualitative different behavior in case of a phase transition.\\[10mm]
\end{abstract}


\begin{multicols}{2}

\section{Introduction}

The azimuthal asymmetries in momentum spectra usually referred to as
directed and elliptic flow are large in nuclear collisions at
intermediate energies \cite{Bevalac} which allows experimental
reconstruction of the reaction plane. Also in relativistic nuclear
collisions at the AGS \cite{AGS} and SPS \cite{NA49v2}
anisotropic flow has been found in semi-central collisions.  Likewise,
HBT analyses has successfully been made, however, so far without
simultaneous reconstruction of the reaction plane. The statistics may
permit this in the near future and the prospects looks good for the
RHIC and LHC colliders, where the multiplicities are much higher and
the number of pairs entering HBT analyses grow with the multiplicity
squared. Particle interferometry (HBT \cite{HBT}) may also show
source asymmetries \cite{Voloshin,Wiedemann} spatially and complement
the momentum space information of flow.
Earlier studies of directed and elliptic flow by hydrodynamics
\cite{Ollitrault} as well as by the cascade code RQMD \cite{RQMD,Xu}
has been compared to AGS and SPS data with some success.  The data
shows interesting transverse momentum and rapidity dependences for
pions and protons that cannot be explained by these models.

The aim of this work is to return to the basic physics that leads to
azimuthal anisotropies in semi-central collisions in both coordinate
and momentum space which are relevant for the HBT radii and anisotropic
flow respectively. We shall derive simple analytical formulas for
elliptic flow and HBT radii in both the collisionless and hydrodynamic
limits relevant for peripheral and near-central collisions
respectively. 
A combined analysis of HBT and anisotropic flow yields
detailed information on the particle source at freeze-out as well as
the expansion from the initial collision and up to freeze-out.
General properties as the initial geometry, shadowing effects,
expansion and freeze-out can be determined through detailed measurements
on source sizes, deformations, lifetimes, duration of emission, opacities,
transverse and elliptic flow.
We discuss the effects of a phase transition occurring
when the energy densities become high enough to
form a quark-gluon plasma at some semi-centrality. 
The centrality, $E_T$ or $dN/dy$ dependence of a
number of measurable quantities as the ratio of the elliptic flow to
the deformation, duration of emission, opacity, etc. may display
an interesting behavior.

The paper is organized as follows. First we describe the geometry of
semicentral relativistic nuclear collisions and discuss the azimuthal
asymmetries and deformations of the overlap zone in sec. II. In
sec. III we calculate elliptic flow in the collisionless and
hydrodynamic limits relevant for peripheral and near central
collisions respectively and compare to recent SPS data \cite{NA49v2}.
HBT radii are calculated for deformed sources in sec. IV.  In sec. V
we discuss the qualitative behavior of a number of physical quantities
from elliptic flow to HBT radii - with and without a phase transition
to a quark-gluon plasma. Finally a summary is given.

\section{Geometry of Semi-central Collisions}

In non-central nuclear collisions the reaction plane breaks azimuthal
symmetry.  The asymmetry decreases with centrality and vanish for the
very central collisions (see Fig. 1). 

We will use the term {\it centrality} as impact parameter
$b$ in the collision. It is not a directly measurable quantity but is
closely correlated to the transverse energy produced $E_T$, the
measured energy in the zero degree calorimeter and the total particle
rapidity density $dN/dy$. The latter is again approximately proportional
to the number of participating nucleons
\be
 N_{part} = \int_{overlap} \left[ \rho({\bf r}+\frac{\bf b}{2})
  + \rho({\bf r}-\frac{\bf b}{2}) \right] d^3r \,. \label{Npart}
\ee
The number of participants for colliding two spherical nuclei of radius
$R$ is shown in Fig. 2 as function of impact parameter.

The standard geometry has the $\rz$-direction along the longitudinal
or beam axis and the $\rx$-direction along the impact parameter ${\bf
b}$.  Thus ({\rm x,z}) constitutes the {\it reaction plane} and
$(\rx,\ry)$ the transverse direction with $\ry$ perpendicular to the
reaction plane. It is convenient --- also for comparison to HBT
analyses --- to employ gaussian parametrizations for anisotropic
sources in both transverse directions
\bea
  S_\perp({\rm x,y}) = \frac{1}{2\pi R_xR_y}
  \exp\left( - \frac{\rx^2}{2R_x^2} -  \frac{\ry^2}{2R_y^2}\right)
   \,. \label{gaussian}
\eea
Here, the gaussian transverse sizes $R_x$ and $R_y$ increase with time
as the source expands after the collision.
The initial transverse radii, i.e., the rms radii of the overlap zone
weighted with nuclear thickness functions, are shown in
Fig. 2.
For HBT the relevant radii are, however, those at freeze-out which
are larger due to expansion as discussed below.

The azimuthal asymmetry or ``deformation'' of the source 
can be defined as the relative difference between the gaussian radii squared
\bea
  \delta \equiv \frac{R_y^2-R_x^2}{R_y^2+R_x^2} \,. \label{delta}
\eea
A simple estimate can be obtained from the full transverse extent of the
initial overlap of two nuclei of mass number $A$ and 
radius $R_A\simeq 1.2A^{1/3}$ colliding with impact parameter
$b$ (see Fig. 1). They are: 
$R_x=R_A-b/2$ and $R_y=\sqrt{R_A^2-b^2/4}$, 
and the corresponding deformation is
\bea
  \delta = \frac{b}{2R_A} \,. \label{deltaa}
\eea
The rms radii of the nuclear overlap zone weighted with longitudinal
thicknesses results in deformations (see Fig. 2) that are slightly smaller
than Eq. (\ref{deltaa}) 
at semicentral collisions. However, as the source expands the
deformation $\delta\simeq(R_y-R_x)/(R_y+R_x)$
decreases for two reasons. 

Firstly, the expansion increase
$(R_x+R_y)$ as is found in HBT and Coulomb
analyses  \cite{Barz}. 
Secondly, $(R_y-R_x)$
decrease because the average velocities are larger in the $x$- than
$y$-direction.  The latter is a consequence of the experimentally
measured positive elliptic flow ($v_2>0$ in Eq. (\ref{v})) in
relativistic nuclear collisions where shadowing is minor
\cite{AGS,NA49v2}.
Measuring the decrease of the deformation with centrality will reveal 
important information on the expansion up to freeze-out. 
For very peripheral collisions, where only a single nucleon-nucleon
collision occurs, the source must be azimuthally symmetric, i.e., the
deformation must vanish and therefore Eq. (4), which assumes continuous
densities, breaks down.

\section{Elliptic Flow}

The reaction
plane, which breaks azimuthal symmetry, has been successfully determined
in  non-central heavy ion collisions from intermediate 
up to ultrarelativistic energies \cite{NA49v2}. 
The particle spectra are expanded in harmonics of the azimuthal
angle $\phi$ event-by-event \cite{Ollitrault}
\bea
  E\frac{dN}{d^3p} &=& \frac{dN}{p_tdp_tdyd\phi} 
  = \int d^4x n(x,p) \,  \nonumber\\
  &=&\, \frac{dN}{2\pi p_tdp_tdy}
  \times
   [1+2v_1\cos(\phi-\phi_R)  \nonumber\\
  && \quad\quad +2v_2\cos2(\phi-\phi_R)\;+... ]  \,, \label{v}
\eea
where $n(x,p)$ is the particle distribution function in space
and time $x={\bf r},t)$, and $\phi_R$ is
the azimuthal angle of the reaction plane. Assuming that the experimental
uncertainties in event plane reconstruction can be corrected for, each
event can be rotated such that $\phi_R=0$. The asymmetry decreases
with centrality (see Fig. 1+2 and Eq. (\ref{deltaa})) and vanish for
the very central collisions which are cylindrical symmetric.  The
expansion parameters are $v_1$ for {\it directed} flow and $v_2$ for
{\it elliptic} flow.

At Bevalac energies where the directed flow in semi-central collisions of
heavy nuclei is up to 20 degrees
\cite{Bevalac} with respect to the beam line. The directed flow
decrease with energy and is only $\sim 1/20$ degree at SPS energies
\cite{NA49v2}.  At AGS and SPS energies there are now very accurate
measurements for protons, positive and negatively charged pions as
function of rapidity and transverse momentum.
\cite{AGS,NA49v2} Both pions and protons are found to have
elliptic flow in the reaction plane at AGS \cite{AGS} and SPS
\cite{NA49v2} energies, i.e., $v_2>0$ for all rapidities and
transverse momenta.  At SPS $v_2\la 0.1$ for protons but smaller for
pions as well as for protons at AGS.  We refer to \cite{Ollitrault} for
recent results on flow in relativistic heavy ion collisions.
 
At collision energies below a few GeV the nuclei shadow the collision
region in the reaction plane. Consequently, the flow is ``squeezed out''
in the ${\rm y}$-direction, i.e., $v_2<0$. At
ultrarelativistic energies Lorentz contraction of the nuclei reduce
the amount of shadowing at midrapidities and one finds that $v_2>0$.
We will in the following restrict ourselves to ultrarelativistic
energies and midrapidities where effects of shadowing are minor.

\subsection{Collisionless limit/Peripheral collisions}

 In peripheral collisions the nuclear overlap zone is small and at
relativistic energies the expansion is fast. Produced particles can
therefore escape frow the collision zone almost without interacting
with the other particles, i.e., the system is close to free streaming
and the collisionless limit. We can then calculate the first order
correction to free streaming from particle collisions.  This
approximation is valid when the particle mean free paths
$\lambda_{mfp}\simeq (\sigma\rho)^{-1}$ are larger than the overlap
zone $R_{x,y}(b)$.  For pions and protons $\lambda_{mfp}$ of order a
few $fm$'s for particle densities of order nuclear saturation
densities, i.e., comparable to source sizes in semi-central
collisions.  If we assume that particles initially are produced
azimuthally symmetric in momentum space but not in coordinate space,
these subsequent interactions with comovers
will produce an azimuthally asymmetric
momentum distribution because the source is azimuthally asymmetric
spatially.

We shall not attempt to describe the initial hard nucleon-nucleon
collisions, the fragmentation or particle production.
Our starting point is some initial conditions at formation time
$\tau_0\sim 1$ fm/c and consider the subsequent scatterings between
comovers described by
the Boltzmann or Nordheim \cite{Nordheim}
equation
\bea
 &&\left( \frac{\partial}{\partial t} + {\bf v}_p\cdot \nabla \right) n_1
 = \int\frac{d^3p_2}{(2\pi)^3} d\Omega \, v_{12}\frac{d\sigma}{d\Omega}\,
 \nonumber\\
 &\times&\left[ n_3n_4(1\pm n_1)(1\pm n_2) - n_1n_2(1\pm n_3)(1\pm n_4) \right]
 \,. \label{BE}
\eea
Here, $n_i=n(r_i,p_i,t)$ is the particle distribution function and
$d\sigma/d\Omega$ is the cms differential cross section for scattering
particles $1+2\to 3+4$.
Stimulated emission and Pauli Blocking are including by the $\pm$ factors.

To evaluate the effect of collisions on the distribution function in
(\ref{BE}) we assume
Bjorken scaling $v_z=z/t$ initially at invariant time $\tau_0$.
As usual, the space-time
rapidity is $\eta=(1/2)\log((t+z)/(t-z))$ and $\tau=\sqrt{t^2-z^2}$ is the
invariant time.
 The transverse particle density distribution is approximated by simple 
gaussians as in Eq. (\ref{gaussian}).
The free streaming distribution is then at later times 
\bea
   n(x,p) = f_0(p_\perp,p_z')\,
   S_\perp({\bf r}_\perp-{\bf v}_\perp(\tau-\tau_0)) \,. \label{fs}
\eea
Here, ${\bf r}_\perp=(\rx,\ry)$ is the transverse radius,
${\bf v}_\perp=(v_x,v_y)$ the transverse velocity.
The local particle momentum distribution, $f_0({\bf p})$, is discussed
in detail in Appendix A; in the final result it can be rewritten
in terms of final measured momentum distributions, $dN/d^2p_t$.
The scaled longitudinal momentum
\bea
  p_z'=\frac{\tau}{\tau_0}m_\perp\sinh(\eta-y) \,, 
\eea
takes the free streaming longitudinally into account\cite{Baym}. 
Transverse expansion
of the source is incorporated by the Galilei transformations of
the transverse coordinates $(\rx,\ry)$.
Note, that the momentum distribution of produced particles is 
azimuthally symmetric at initial production time $\tau_0$.
The detailed form of $f_0$ will not be necessary for the
evaluation of elliptic flow or HBT radii. It is sufficient to
know the rapidity density $dN_j/dy$ of the scattering particles $j$.

In the collisionless limit we can now insert the free streaming distribution
function in the collision term in order to calculate the first order
correction to the distribution function from (\ref{BE}). This correction
provides the deviation from cylindrical symmetry and directly leads to elliptic
flow, $v_2^i$, for particle species $i=\pi,p,K,...$, as
is evaluated in detail in Appendix A
\bea
 v_2^i = \frac{\delta}{16\pi R_xR_y}
  \sum_j \langle v_{ij}\sigma_{tr}^{ij} \rangle \frac{dN_j}{dy}
  \frac{v_{i\perp}^2}{{v_{i\perp}^2+\langle v_{j,\perp}^2\rangle}}
     \,  \,. \label{v2c}
\eea
Here, $v_i$ is the particle, $v_j$ the scatterer and $v_{ij}$ the relative
velocity.  The averages $\langle ..\rangle$
refers to averaging over scatterer momenta $p_j$.
Since it is the momentum transfer in collisions, that
deforms the particle momentum distribution around cylindrical
symmetry, it is naturally the {\it momentum transport cross section}
$\sigma_{tr}^{ij}$ that enters.  The elliptic flow parameter
is proportional to the deformation and neccessarily vanishes for an
azimuthally symmetric source.

The other important factors in the elliptic flow are 
\bea
  \tilde{\sigma}_{ij} \equiv 
  \frac{dN_j}{dy} \frac{\langle v_{ij}\sigma_{ij}\rangle}{\pi R_xR_y} \,.
\eea
It is the cross section times a transverse
particle density and thus describes the effective ``opacity'' of the
source for particles {\it i} scattering with particles {\it j}.

It is remarkable that the initial time $\tau_0$ does not enter in
(\ref{v2c}).  The $\tau_0$-dependence cancels to first approximation 
as explained
in appendix A, due to a compensation between the densities decreasing
with expansion time and that scatterings only lead to asymmetries as
they pass through the source with time.  The physical reason for the
``late time'' dominance in elliptic flow is that a particle has to
travel a distance $\sim R$ before it feels the source deformation,
i.e., whether it moves in the $\rx$-direction, or in the $\ry-$
direction which differs by a distance $R_y-R_x\simeq\delta R$. This
also explains why the elliptic flow is proportional to
$v_{i\perp}^2$. The slow particles do not travel far enough to feel
the deformation before the source has expanded - by velocity
$v_{ij,\perp}$.

The centrality dependence of the elliptic flow can be estimated from
Eq. (\ref{v2c}). In high energy nuclear collisions the total
multiplicity $dN/dy$ scales approximately with the number of
participating nucleons, Eq. (\ref{Npart}) which is shown in Fig. 2 as
function of impact parameter together with $R_x(b)$, $R_y(b)$ and
$\delta(b)$. The resulting centrality dependence of $v_2$ from
Eq. (\ref{v2c}), assuming that $dN_j/dy$ also are proportional to the
number of participants, is shown in Fig. 2 as well. It vanishes for
central and grazing collisions, has a maximum for peripheral
collisions $b\simeq 1.5R$ and decreases almost linearly with
increasing centrality.  A very similar behavior was found in the
hydrodynamic calculations of Ollitrault \cite{Ollitrault}.

\subsection{Hydrodynamic limit/Semi-central collisions}

In semi-central collisions at relativistic energies, where large particle
densities are produced, rescatterings are abundant and the
hydrodynamic limit (see, e.g., \cite{Ollitrault}) is better than
the collisionless --- at least up to freeze-out.

We will here attempt to extract some general features in the
hydrodynamic limit and derive an approximate analytical formula for
the elliptic flow.  Instead of introducing two independent transverse
flow velocities $(u_x,u_y)$ we relate the transverse flow to the
spatial deformation by the plausible assumption that the transverse
flow is {\it equipotential, i.e., perpendicular to and with constant
magnitude on equi-density surfaces}.  This assumption is obvious for
cylindrical symmetric sources.  Also for a very deformed source, which
appears as a slab in one transverse direction, the flow is
perpendicular to most of the freeze-out surface.  For a rectangular
source the flow is initially also perpendicular to the surface in all
directions but as the source expands deviations may occur at the
corners.  Concerning the constant magnitude of the transverse flow on
the freeze-out surface, that may be justified for deformed sources by
the following observation.  In the Riemann solution to 1D
hydrodynamics, the flow velocity only depends on time and the relative
distance from the initial surface, i.e., $(r-R)$. As shown in
\cite{Bhydro} the Riemann solution is also a very good approximation to
the cylindrical case with longitudinal Bjorken scaling with an
additional scaling factor of $(\tau_0/\tau)^{c_s^2}$ for
temperatures. For the slab or the rectangular source, the magnitude
flow velocities will therefore only depend on the relative distances
from the initial transverse radii, i.e., $(\rx-R_x)$ or $(\ry-R_y)$.
How good this approximation is in general for near central
relativistic heavy ion collisions should, of course, be checked by
solving full 3D hydrodynamics.

With the simplifying approximation of equipotential flow
we can calculate the resulting
particle spectrum for a locally thermal
distribution $e^{p\cdot u/T}$ with longitudinal Bjorken flow, 
\bea
   u=(\gamma\cosh(\eta),\gamma\sinh(\eta),{\bf u}_\perp) , \label{u}
\eea
where $\gamma=\sqrt{1+u_\perp^2}$.
The transverse extent is parametrized by (\ref{gaussian}).
With these assumptions the resulting momentum distribution can be
found by integrating over the deformed equipotential surfaces 
as described in Appendix B. The asymmetric term gives the elliptic flow
\bea
  v_2 =  \left\{ \begin{array}{lrl}
    \frac{1}{4} \frac{p_\perp^2 \langle u_\perp^2\rangle}{T^2} \delta 
    & , & p_\perp u_\perp\ll T \\
    \frac{1}{2} (\frac{p_\perp \langle u_\perp\rangle}{T}-1) \delta 
      & , & p_\perp u_\perp\gg T \end{array} \right\} \,.   \label{v2h}
\eea  
It is expected that the elliptic flow is proportional to the
deformation which vanishes for central collisions. The relevant
deformation is now at freeze-out where the radii which
may differ from the initial transverse radii as discussed above. 

The elliptic flow of Eqs. (\ref{v2h}) depends only indirectly on the
unknown quantities as the initial densities, equation of state,
expansion, etc. through the average transverse flow, temperature and
deformation. The transverse flow and temperature can be extracted
independently from transverse flow analyses by looking at the particle
mass dependence of average $p_\perp$ as suggested in \cite{Kataja} or
by measuring the {\it apparent temperatures} \cite{NA44slopes}, i.e.,
the inverse of the $m_\perp$ slopes.  As shown in Appendix C the
apparent temperature obtained from exponential fits to transverse mass
spectra of massive particles is
\bea
  T_{app}\simeq T + \frac{1}{2} m\langle u_\perp^2\rangle \,, \label{Tapp}
\eea
for large particle masses $m$ and small transverse flow.
Experimentally, the apparent temperatures scales approximately linearly
with the masses of the pion, kaon, proton, and deuterium in central
$S+S$ and $Pb+Pb$ collisions at AGS and CERN \cite{NA44slopes} energies.
From the experimental slopes $2dT/dm$ we thus obtain
$\langle u_\perp^2\rangle \simeq 0.15c^2$ and $0.3c^2$ 
for central $S+S$ and $Pb+Pb$ collisions at SPS energies.

It is instructive to compare to 
another type of hydrodynamic flow parametrized by two transverse flow
velocities 
\bea
 {\bf u}_\perp=(u_x x/R_x,u_y y/R_y) \, .\label{uxy}
\eea 
The resulting particle distribution can be calculated along the lines of
Appendix B. The resulting elliptic flow is for small transverse momenta
\bea
  v_2 =  \left\{ \begin{array}{lrl}
  \frac{1}{16} \frac{p_\perp^2}{T^2} 
    (\langle u_x^2\rangle-\langle u_y^2\rangle)
    & , & p_\perp u_\perp\ll T \\
    \frac{1}{8} (\frac{p_\perp \langle u_\perp\rangle}{T}-1) 
   \frac{\langle u_x^2\rangle-\langle u_y^2\rangle}{u_\perp^2} 
      & , & p_\perp u_\perp\gg T \end{array} \right\} \,.   \label{v2xy}
\eea  
Comparing Eq. (\ref{v2xy}) to the elliptic flow data for protons of Fig. 3 
we can estimate the difference 
$\langle u_x^2\rangle-\langle u_y^2\rangle \simeq 0.015$, i.e., an
order of magnitude smaller than the average
transverse flow $u_\perp^2\simeq(u_x^2+u_y^2)/2\simeq 0.15$ that was estimated
from Eq. (\ref{Tapp}).
The elliptic flow of Eq. (\ref{v2xy}) is not related to the
spatial deformation as was the equipotential flow and therefore its
predictive power is less.

\subsection{Comparison to elliptic flow at SPS energies}

The elliptic flow as estimated by the collisionless and hydrodynamic
limits may be compared to NA49 data at SPS energies \cite{NA49v2} taken
for semicentral $Pb+Pb$ collisions ($b\simeq R\simeq 7fm$).  The
$p_\perp$ dependence of $v_2$ for pions and protons is measured at
forward rapidities $4\le y\le5$.  We can take the free cross sections
and ignore medium effects since the densities are small at late times
when rescatterings lead to anisotropies (see Appendix A).  The relevant
scattering cross section $\sigma^{ij}$ is that where nucleons act as
scatterers because they are heavier than e.g. pions and thus deflect
more.  Furthermore, typical scattering cross sections $\sigma^{\pi N}$
are larger than $\sigma^{\pi\pi}\simeq 10$mb \cite{Bertsch}.  The
former is dominated by the $\Delta$-resonance. Averaging of scatterer
momenta we estimate $\sigma^{\pi N}\simeq 30$mb and $\sigma^{NN}\simeq
40$mb.  The transport cross sections are smaller because forward
scattering without momentum loss should not be included. For an
estimate we simply take the transport cross section as half of the
total cross section.  The relevant rapidity density is that of
nucleons which we take from the NA49 experiment $dN_N/dy\simeq
30$. From Fig. 2 we obtain $R_x\simeq 2fm$, $R_y\simeq 3fm$ and
$\delta\simeq 0.4$ for $b=R$.  The resulting elliptic flow is shown in
Fig. 3 for pions and protons.  The estimates in the collisionless
limit of Eq.(\ref{v2c}) give a reasonable description of both the
magnitude and $p_\perp$-dependence for both pions and protons.
Furthermore, the measured elliptic flow is largest at midrapidities and
is approximately proportional to $dN/dy$ as predicted by
Eq.(\ref{v2c}).

In the hydrodynamic limit transverse flow couples to $p_\perp$ and
elliptic flow is therefore expected to be the same for all particles
at the same $p_\perp$. The elliptic flow can differ if various
particle species freeze-out at different temperatures or resonance
decays affect the final distributions. The elliptic flow is shown in
Fig. 3 for equipotential flow, Eq. (\ref{v2h}), with $T\simeq 160$
MeV and $\langle u_\perp^2\rangle\simeq 0.15c^2$ as estimated above.
An excellent fit to the magnitude and $p_\perp$ dependence of proton
elliptic flow can be obtained with a deformation at freeze-out of
$\delta\simeq0.07$. This value for the deformation is a factor $\sim6$
smaller than the initial deformation of the collision zone.  It is
expected to be smaller due to expansion between initial collision and
freeze-out but a full 3+1 dimensional hydrodynamic calculation is
needed in order to chech the magnitude of the deformation as well as
the validity of equipotential flow.  The deformation at freeze-out can
in principle be extracted from HBT analyses as shown in the following
section and would thus be an important independent check.

 The pion elliptic flow cannot be explained in the hydrodynamic limit
as it differs from proton flow and does not have the right $p_\perp$
dependence.  It is curious that the magnitude of the pion elliptic
flow lies between the collisionless and hydrodynamic limit.  This may
indicate that semi-central $Pb+Pb$ collisions at SPS energies lies
between these two limits as one would expect since the particle mean
free paths are a few $fermi$'s --- comparable to the transverse source
sizes for $b=R_{Pb}\simeq 7$ fm/c.  Comparing the $p_\perp$-dependence
of elliptic flow for pions and protons would reveal the change from
collisionless expansion to hydrodynamic flow. In the collisionless
limit the pion and proton elliptic flow differ whereas in the
hydrodynamic limit they should become the same as shown in Fig. 3.

Calculations with the RQMD and VENUS models (see \cite{NA49v2,Xu} for
detailed comparisons) give an almost rapidity independent
elliptic flow which underpredicts the SPS data by a factor $\sim2-4$ at
midrapidity. The smaller elliptic flow in RQMD at midrapidities is
attributed to ``preequilibrium softening'' \cite{RQMD}. Hydrodynamic
models results in corresponding values for $v_2$ which are $\sim4$
times larger for an ideal $(c_s^2=1/3)$ pion gas and $\sim2$ times
larger when a first order phase transition at $T_c\sim150$MeV is
included \cite{Ollitrault}.

It is curious that the elliptic flow is less than a few percent in
relativistic nuclear collisions in comparison to the initial spatial
deformations which are of order $\delta\simeq50\%$ in semi-central collisions.
The reason can be understood in both the collisionless and hydrodynamic
limits due to the limited time and distances that particles have to
rescatter and develop asymmetric collective flow before they freeze-out.
In other words, the scatterings that lead to asymmetric flow occur at
late times where the source and scatterers have already expanded 
and reduced the initial deformation.

The elliptic flow at RHIC and LHC energies would be large if one
simply extrapolates from intermediate, AGS and SPS energies
\cite{NA49v2} However, at the intermediate energies shadowing is
responsible for a negative $v_2$ (squeeze-out) and is still felt at
AGS energies.  At SPS energies it may also affect the forward rapidity
data $4<y<5$ where the projectile nucleus is less Lorentz contracted
and thus shadow.  But at central rapidities at SPS and higher energies
shadowing is minor due to the strong Lorentz contraction of nuclear
sizes.  Therefore, $v_2$ is unaffected by shadowing from SPS and up in
energy, i.e., constant rather than increase as given by a naive
extrapolation from intermediate, AGS and SPS energies.  However, since
$dN/dy$ and probably also the transverse flow is larger at RHIC and
LHC energies, we may expect stronger elliptic flow. On the other hand,
the expansion reduces the deformation and thus also the elliptic flow.

\section{HBT}

A brief description of interferometry will be given describing how to
calculate correlation functions and HBT radii from a given source.
The HBT radii will then be calculated for deformed sources -
transparent as well as opaque. 
A brief description of results were presented in \cite{HBD}.
A comparison is given and the
advantages of a combined HBT and elliptic flow analysis is discussed.

 Particle interferometry was invented
by Hanbury-Brown \& Twiss (HBT) for stellar size determination
\cite{HBT} and is now employed in nuclear collisions
\cite{GKW,Pratt,Csorgo,Heinz,HH,NA49,NA44}.  It is a very powerful
method to determine the 3-dimensional source sizes, life-times,
duration of emission, flow, etc. of pions, kaons, etc. at
freeze-out. Since the number of pairs grow with the multiplicity per
event squared the HBT method will become even better at RHIC and LHC
colliders where the multiplicity will be even higher. 

The standard HBT method for calculating the Bose-Einstein correlation function
from the interference of two identical particles is now briefly discussed.
For a source of size $R$ we consider two particles emitted a distance
$\sim R$ apart with relative momentum ${\bf q}=({\bf k}_1-{\bf k}_2)$
and average momentum, ${\bf K}=({\bf k}_1+{\bf k}_2)/2$. Typical heavy
ion sources in nuclear collisions are of size $R\sim5$ fm, so that
interference occurs predominantly when 
$q\raisebox{-.5ex}{$\stackrel{<}{\sim}$}\hbar/R\sim 40$
MeV/c. Since typical particle momenta are $k_i\ga K\sim 300$ MeV/c,
the interfering particles travel almost parallel, i.e.,
$k_1\simeq k_2\simeq K\gg q$.  The correlation function due to
Bose-Einstein interference of identical spin zero bosons as
$\pi^\pm\pi^\pm$, $K^\pm K^\pm$, etc.
from an incoherent source is (see, e.g., \cite{Heinz})
\begin{equation}
  C_2({\bf q},{\bf K})=1\;\pm\; \left|\frac{\int d^4x\;S(x,{\bf K})\;e^{iqx}}
   {\int d^4x\;S(x,{\bf K})}\right|^2 \,, \label{C}
\end{equation}
where $S(x,{\bf K})$ is the source distribution 
function describing the phase space density of the
emitting source. 

Experimentally the correlation functions are often
parametrized by the gaussian form
\bea
  C_2(q_s,q_o,q_l)&=&1+\lambda\exp[
           - q_s^2R_s^2-q_o^2R_o^2-q_l^2R_l^2  \nonumber\\
  &&  -2q_oq_sR_{os}^2 
            -2q_oq_lR_{ol}^2 ]\;.   \label{Cexp}
\eea
Here, ${\bf q}={\bf k}_1-{\bf k}_2=(q_s,q_o,q_l)$ is the relative
momentum between the two particles and $R_i,i=s,o,l,os,ol$ the
corresponding sideward, outward, longitudinal, out-sideward and
out-long HBT radii respectively.  We have suppressed the ${\bf K}$
dependence.  We will employ the standard geometry, where the {\it
longitudinal} direction is along the beam axis, the {\it outward}
direction is along ${\bf K}_\perp$, and the {\it sideward} axis is
perpendicular to these.  Usually, each pair of particles is lorentz
boosted longitudinal to the system where their rapidity vanishes,
$y=0$. Their average momentum ${\bf K}$ is then perpendicular to the
beam axis and is chosen as the outward direction. In this system the
pair velocity \mbox{\boldmath $\beta_K$}=${\bf K}/E_K$ points in the
outward direction with $\beta_o=p_\perp/m_\perp$, where
$m_\perp=\sqrt{m^2+p_\perp^2}$ is the transverse mass, and the
out-longitudinal coupling $R_{ol}$ vanish at midrapidity \cite{Heinz}. 
Also $R_{os}$ vanishes for a cylindrically symmetric source or if
the azimuthal angle of the reaction plane is not determined and
therefore averaged over --- as has been the case experimentally so far.
The reduction factor $\lambda$ in Eq. (\ref{Cexp})
may be due to long lived resonances \cite{Csorgo,HH,Wiedemann},
coherence effects, incorrect Coulomb corrections or other effects. It
is $\lambda\sim 0.5$ for pions and $\lambda\sim 0.9$ for kaons.

 The Bose-Einstein correlation function can now be calculated for a
deformed source.  Let us first investigate {\it transparent} sources
assuming that it is equally likely that a particle
arrives at the detector from the front side of the
source as from the back side, i.e., rescatterings through the source 
and opacities are ignored. 
As in \cite{Wiedemann}, we parametrize the transverse and temporal
extent by gaussians
\bea
   S(x,K) \sim S_\perp(\rx,\ry) \exp[
         -\frac{(\tau-\tau_f)^2}{2\delta\tau^2}] \, e^{p\cdot u/T} 
   \,, \label{S}
\eea
with longitudinal Bjorken flow,
$u=(\cosh\eta,0,0,\sinh\eta)$, 
as in Eq (\ref{u}) without transverse flow. Effects of 
transverse flow will be discussed below.
The transverse radii $R_x,R_y$ are the gaussian radii at freeze-out, 
$\tau_f$ is the freeze-out time and $\delta\tau$ the duration of emission.

In order to calculate the correlation function of (\ref{C}) the
gaussian approximations is employed (see, e.g., \cite{Csorgo,Heinz})
which results in a correlation function of the form given in
Eq. (\ref{Cexp}). Inserting the source (\ref{S}) in Eq. (\ref{C}) and
Fourier transforming we obtain the correlation function. 
The HBT radii are then obtained by comparing to the 
experimental correlation function of Eq.(\ref{Cexp}). 

It is convenient to introduce the source average 
of a quantity ${\cal O}$ defined by
\begin{equation}
 \langle{\cal O}\rangle\equiv
   \frac{\int d^4x\;S(x,{\bf K}){\cal O}}{\int d^4x\;S(x,{\bf K})}\;. 
 \label{O}
\end{equation} 
With $qx\simeq{\bf q\cdot x}-{\bf q\cdot}$\mbox{\boldmath $\beta$}$_K\,t$
one can, by expanding Eq. (\ref{C}) 
to second order in $q_iR_i$ and compare to
Eq. (\ref{Cexp}), find the HBT radii \cite{Heinz}
\bea
  R_i^2 &=& \langle (x_i-\beta_i\;t)^2\rangle-\langle x_i-\beta_i\;t\rangle^2
 \,, \label{Ri} \\
  R_{ij}^2 &=& \langle (x_i-\beta_i\;t)(x_j-\beta_j\;t)\rangle
 -\langle x_i-\beta_i\;t\rangle\langle x_j-\beta_j\;t\rangle    \,,
 \nonumber\\
\eea
with {\it i,j=s,o,l}.  The HBT radii are a measure for the
fluctuations \cite{Opaque}, variance or ``length of homogeneity''
\cite{Sinyukov} of $(x_i-\beta_it)$ over the source emission function
$S(x,K)$.  One should notice that the coordinates $(x_o,x_s)$ are
rotated with respect to the $(\rx,\ry)$ reaction plane (see Fig. 1) by
the azimuthal angle $\phi$ between the transverse momentum $p_\perp$
and the reaction plane. The beam axis is the same $x_l=\rz$.  In the
local cms frame ($y=0$) the pair velocity is $\beta_o=\beta_\perp$
whereas $\beta_s=\beta_l=0$.  All $R_{ij}$ vanish in this frame for
cylindrical symmetric sources and for an azimuthally asymmetric source
only $R_{os}$ is nonvanishing.

For transparent sources the azimuthal dependence of
the HBT radii has been calculated in detail by Wiedemann
\cite{Wiedemann}. In the longitudinal center-of-mass system
of the pair $(y=0)$, the HBT radii are
\bea
   R_s^2 &=& R^2\left[1 + \delta\cos(2\phi)\right] \,, \label{Rst} \\
   R_o^2 &=& R^2\left[1 - \delta\cos(2\phi)\right] 
             \,+ \beta_o^2\delta\tau^2  \,,   \label{Rot} \\
   R_{os}^2 &=& R^2\delta \sin(2\phi) \,, \label{Rost}\\
   R_l^2 &=& \frac{T}{m_\perp} (\tau_f^2+\delta\tau^2) \,, \label{Rl}
\eea
where $R^2=(R_x^2+R_y^2)/2$ is the average of the source radii
squared. One notices the characteristic modulation $\cos(2\phi)$ due
to the rotation of the axes $R_{o,s}$ with respect to the reaction plane.
Near target and
projectile rapidities the directed flow is appreciable and leads to
$\cos\phi$ terms in Eqs. (\ref{Rst}-\ref{Rot}) \cite{Ollitrault,Wiedemann}.
The out- and sideward HBT radii show a
characteristic modulation as function of azimuthal angle with
amplitude of same magnitude but opposite sign. 
Measuring the amplitude modulation of $R_{s,o,os}$ determines five quantities
and thus overdetermines the three source parameters, namely 
the source size  $R$, deformation $\delta$ and 
and duration of emission $\delta\tau$.

Next we consider {\it opaque} sources. In relativistic heavy ion
collisions source sizes and densities are large and one would expect
rescatterings. As a result particles are predominantly emitted near the
surface and arrive from the (front) side of the source facing
towards the detector.  In \cite{Opaque} it was found that for
opaque sources, where mean free paths are smaller than source sizes,
$\lambda_{mfp}\la R$, the sideward HBT radius increase whereas the
outward is significantly reduced simply because the surface emission
region is smaller than the whole source.  As in \cite{Opaque,Tomasik}
Glauber absorption is introduced by adding an absorption factor 
\bea
  S_{abs}(x,K) \sim \exp(-\int_x \sigma\rho(x')dx') \,,
\eea
where $\sigma$ is the interaction cross section,
$\rho$ the density of scatterers and the integral runs along the
particle trajectory from source point $x$ to the detector. Defining
the mean free path as $\lambda_{mfp}=(\sigma\rho(0))^{-1}$, where
$\rho(0)$ is the central density, the source is opaque when
$\lambda_{mfp}\ll R$ and transparent when $\lambda_{mfp}\gg R$.
Calculating the correlation function for an opaque source from
Eq. (\ref{C}) and comparing to the definition of the HBT radii in
Eq. (\ref{Cexp}), one generally obtains for small deformations,
\bea
   R_s^2 &=& g_sR^2\left[1 + \delta\cos(2\phi)\right] \,, \label{Rso} \\
   R_o^2 &=& g_oR^2\left[1 - \delta\cos(2\phi)\right] 
             \,+ \beta_o^2\delta\tau^2  \,,           \label{Roo} \\
   R_{os}^2 &=& g_{os} R^2\delta \sin(2\phi) \,, \label{Roso}
\eea
and $R_l^2$ is unchanged from Eq. (\ref{Rl}).
Here $g_{o,s,os}$ are model dependent factors that are functions of
opacity but independent of the deformation and can be calculated as
for cylindrical symmetric sources.  For a gaussian source
($\rho\propto S_\perp$), which is moderately opaque
$(\lambda_{mfp}/R=1$), a numerical calculation gives $g_s\simeq 1.4$
and $g_o\simeq 0.9$ (see also \cite{Tomasik}). 
For a disk source with the same rms
transverse radius as the gaussian radius (i.e., $R_{disk}=2R$), 
that emits like a
black body (i.e., $\lambda_{mfp}\ll R$), one finds $g_s=4/3$ and
$g_o=4(\frac{2}{3}-(\frac{\pi}{4})^2)\simeq 0.2$ \cite{Opaque}.
Generally, $(g_s-g_o)\ge 0$ and the
difference increases with opacity. Only for a completely transparent
sources is $g_s=g_o$.  In all
cases $g_{os}\simeq g_s$.
In Fig. 4 the HBT radii of
Eqs. (\ref{Rst}-\ref{Roso}) are shown for a near-central collisions
($\delta=0.4$) with a moderate duration of emission
($\beta_o^2\delta\tau^2/g_sR^2=0.5$) for various opacities
$\lambda_{mfp}/R=0.1,0.5,1.0,2.0,\infty$. 
As the opacity increases, $g_o/g_s$ decreases and therefore also
the outward HBT radius and its amplitude.

Comparing the HBT radii from an opaque source
Eqs.(\ref{Rso}-\ref{Roso}) with those of a transparent source
Eqs.(\ref{Rst}-\ref{Rost}), one notices that {\it the amplitudes in
$R_s$ and $R_o$ differ} by the amount $(g_s-g_o)$.
The modulation of the HBT radii with $\phi$
provides five measurable quantities which over-determinates
the four physical quantities describing the source: its size $R$, deformation
$\delta$, opacity $(g_o-g_s)$ and duration of emission $\delta\tau$, at
each impact parameter.  The azimuthal dependence of the HBT radii thus
offers an unique way to determine the opacity of the source as well as
the duration of emission separately.  

Experimentally, HBT analyses have not been combined
with determination of the reaction plane yet. Consequently, the azimuthal
angle $\phi$ is averaged and the information on three of the five measurable
quantities in Eqs. (\ref{Rso}-\ref{Roso}) is lost. From the angular averaged
difference between the out- and sideward HBT radii
\bea
 \langle R_o^2-R_s^2\rangle_\phi = \beta_o^2\delta\tau^2 - (g_s-g_o)R^2 
 \,, \label{Ro-Rs}
\eea
one can only determine the sum of the positive duration of emission
and negative opacity effect.  Experimentally this difference is small;
NA49 \cite{NA49} and NA44 \cite{NA44} data even differ on the
sign. Detailed analyses of the $p_\perp$ dependence of the HBT radii
from NA49 data within opaque sources \cite{Tomasik} indicate that the
sources are transparent or at most moderately opaque. However, the
NA44 data, for which $R_o\la R_s$, requires an opaque source as seen
from Eq. (\ref{Ro-Rs}).  Furthermore, the $p_\perp$ dependence of the
transverse HBT radii change if the source sizes, opacities, and
duration of emission also are $p_\perp$-dependent.

Transverse flow may affect the out- and sideward HBT radii as opacity,
i.e., the factors $g_{o,s}$ may depend on both.  For transparent
sources transverse flow has been studied in
\cite{Wiedemann,Csorgo,Tomasik} through Eq. (\ref{u}) assuming that
the transverse flow scales with transverse distance, ${\bf
u}_\perp\simeq u_0 {\bf r}_\perp/R$. Otherwise the same transparent
gaussian source as in Eq. (\ref{S}) was employed. To lowest order in
the transverse flow both transverse HBT radii decrease by the same
factor $\sim(1+u_0^2m_\perp/T)$ to leading order in $u_0$.  This
transverse flow correction is independent of the source size and
therefore also the deformation. Consequently, spatial deformations
reduce the amplitudes by the same amount in this model for both
transparent (see \cite{Wiedemann}) and opaque sources.  There are,
however, box shape models where the transverse flow reduce $R_o$ more
than $R_s$ \cite{Tomasik}. A similar flow effect is found in
hydrodynamic models also \cite{Rischke} although it is considerably
less than the opacity effect, i.e., when the Cooper-Frye freeze-out
condition \cite{CF} is replaced by the Bugaev freeze-out
\cite{Bugaev}.  Preferrably, a transverse flow analysis of particle
spectra should be performed in order to determine the magnitude of the
flow so that it can be separated from opacity effects. The crucial
question, however, is whether the transverse flow is azimuthally
dependent, i.e., $g_{s,o}$ depend on $\phi$ and thereby change the
amplitudes? We can estimate this effect from the elliptic flow.  For a
simple thermal source (\ref{S}) with transverse flow as in
Eq. (\ref{uxy}), the resulting elliptic flow Eq. (\ref{v2xy}) led to a
very small difference for semicentral $Pb+Pb$ collisions at SPS
energies $\langle u_x^2\rangle-\langle u_y^2\rangle$ less than a few
percent. Therefore, the azimuthally dependence of the flow and its
effect on the amplitude in the HBT radii is also of that order only
which is much less than $\delta\simeq 0.5$ for semi-central collisions.
Both the deformation and elliptic flow decrease with centrality. The
conclusion is that besides opacity also
transverse flow may affect the factors $g_{o,s}$
- but independently of azimuthal angle. Therefore
Eqs. (\ref{Rso}-\ref{Roso}) are still valid and can be used to extract
the duration of emission unambiguously.

It is very important to measure the centrality or impact parameter
dependence of the source sizes, deformation, opacity, emission times
and duration of emission in order to determine how the source change
with initial energy density. If no phase transition takes place one
would expect that source sizes and emission times increase gradually
with centrality whereas the deformation decrease approximately as in
Eq. (\ref{delta}). In peripheral collisions, where source sizes and
densities are small and few rescatterings occur, the source is
transparent and the HBT radii are given by Eqs. (\ref{Rst}-\ref{Rl}).
For near central collisions source sizes and densities are higher
which leads to more rescatterings. Thus the source is more opaque and
the amplitudes should differ. It would be interesting to observe this
gradual change in the amplitudes with centrality. At the same time it
would provide a direct experimental determination whether the source
is transparent or opaque as well as extracting the magnitudes of the
opacity and duration of emission separately.

\section{Effects of a Phase Transition}

If the matter created in an ultra-relativistic heavy ion collision
undergoes a phase transition it may affect the elliptic
flow and the modulation of the HBT radii. The critical energy density
can be overcome either by increasing the centrality, sizes of the
colliding nuclei and/or by
increasing the collision energy as will be possible at the RHIC collider,
$\sqrt{s}\simeq 20\to 200$ GeV.  For energy densities just above the
critical value a mixed phase is created for a first order transition
with zero compressibility, i.e., vanishing sound speed. 
Even for a second order or smooth transition, the  sound speed 
$c_s^2=dP/d\epsilon$
is significantly reduced according to lattice calculations \cite{Lattice}.
The resulting
transverse flow is therefore reduced \cite{Kataja} and should lead to
smaller elliptic flow according to Eq. (\ref{v2h}).  

Unfortunately, the elliptic flow and the modulation of the HBT radii
both scale with the deformation and thus decrease linearly with
centrality and vanish for near central collisions, where the phase
transition might occur. However, by taking the ratio of the elliptic
flow to the amplitude of the HBT radii $v_2/\delta$, the deformation
cancel and the purely geometrical effects are removed.  Only the
collective effects as transverse flow are left to first approximation.
Since the transverse flow has been found to increase with the size of
the collisions system at AGS and SPS energies, we can also expect it
to increase with centrality in the absence of phase transitions.  In
the presense of a phase transition, however, the transverse flow
increase very little with centrality which results in a ``plateau'' in
the average transverse momentum as function of centrality
\cite{vanHove,Kataja}. The almost constant transverse flow for centralities
that produce a mixed phase should therefore according to
Eq. (\ref{v2h}) also show up in the ratio $v_2/\delta$ versus
centrality such that the monotonic increase in transverse flow and
$v_2/\delta$ is replaced at the critical centrality by a plateau as
function of centrality as sketched in Fig. 5.  The plateau extends in
an interval of semi-centrality, $c_1\to c_2$, where the pressure is
constant due to a mixed phase, i.e., vanishing sound speed.

The functional dependence of $v_2/\delta$ would be similar to the 
van Hove prediction of the temperature as function of rapidity density
\cite{vanHove}. Also it is similar to the {\it
caloric curve} as was recently observed in intermediate energy nuclear
collisions \cite{CalCurve}. The caloric curve plots the temperature
versus excitation energy and displays a plateau as was predicted for
the liquid-gas phase transition of nuclear matter \cite{Bondorf}.
The interpretation of the caloric curve is, however, 
debated in the low energy community (see, e.g., \cite{Hauger}).

If a phase transition occurs at some centrality, where energy
densities exceed the critical value, one may also observe sudden
changes in the physical quantities measured in HBT analyses. The
emission time and duration of emission increase drastically in
hydrodynamic calculations \cite{Rischke} leading to very large $R_l$
and $R_o$ (see also Fig. 5). A long lived mixed phase would also emit
particles as a black body and thus the opacity increase and
the opacity factor $g_o/g_s$ decrease as shown in Fig. 5. 
If droplet formation occurs
leading to rapidity fluctuations, one may be able to trigger on such
fluctuations and find smaller longitudinal and sideward HBT radii
\cite{HJ}.  

 Alternatively, one could plot the quantities as function
of collision energy or size of the colliding nuclei
for fixed centrality with a similar qualitative
dependence of the physical quantities with and without a phase
transition.  Furthermore, if an interesting change in these quantities
should occur at some centrality or collision energy, it would also be
most interesting to look for simultaneous $J/\Psi$ suppression \cite{J},
strangeness enhancement, decrease in directed, elliptic or transverse
flow, or other signals from a phase transition.

\section{Summary}

In summary, measuring the reaction plane in relativistic nuclear
collisions, the elliptic flow and HBT radii simultaneously gives a
detailed description of the source at freeze-out as well as how the
source expands in space and time from the initial collision and up to
freeze-out.

The elliptic flow was calculated in the collisionless limit from the
Boltzmann equation with a collision term as well as in the
hydrodynamic limit assuming equipotential flow. The collisionless
limit is expected to be valid for peripheral collision where few
rescatterings are expected. Alternatively, the hydrodynamic limit is
expected to be valid for central collision where many rescatterings
are expected.  For small transverse source deformations the elliptic
flow is proportional to the deformation in both limits. Detailed
cascade and 3+1 dimensional hydrodynamic calculations are clearly
needed of both the radial and elliptic flow as well as HBT radii for
non-central nuclear collisions with and without phase transitions.

The estimates in the collisionless limit of Eq.(\ref{v2c}) give a
reasonable description of both the magnitude and $p_\perp$-dependence
of the elliptic flow measured at SPS energies in $Pb+Pb$ semicentral
collisions for both pions and protons.  Furthermore, the measured
elliptic flow is largest at midrapidies and is approximately
proportional to $dN/dy$ as also predicted by Eq.(\ref{v2c}).  In the
hydrodynamic limit transverse flow couples to $p_\perp$ and elliptic
flow is therefore expected to be the same for all particles at the
same $p_\perp$.  For equipotential flow given by Eq. (\ref{v2h}) can
also describe the proton elliptic flow , however, with a deformation
is a factor $\sim7$ smaller than the initial deformation of the
collision zone. It is expected to be somewhat smaller due to expansion
between initial collision and freeze-out.  The measured elliptic flow for
pions lies between the collisionless and hydrodynamic limit as could
be expected since the particle mean free paths are a few $fermi$'s ---
comparable to the transverse source sizes for $b=R_{Pb}\simeq 7$ fm/c.
Comparing the $p_\perp$-dependence of elliptic flow for pions and
protons reveals distinctively the change from collisionless expansion
to hydrodynamic flow. In the collisionless limit they differ (see
Fig. 3) whereas in the hydrodynamic limit they should become the same. 

The modulation of the HBT radii with azimuthal angle between the
reaction plane and particle transverse momenta can be exploited to
obtain source sizes, deformations, life-times, duration of emission
and opacities separately.  HBT radii provides important
space-time information which complements the momentum space
information from particle spectra on elliptic flow during expansion
and freeze-out.  The HBT radii can distinguish between opaque
sources Eqs.(\ref{Rso}-\ref{Roso}) from transparent source
Eqs.(\ref{Rst}-\ref{Rost}) because {\it the amplitudes in $R_s$ and
$R_o$ differ}. The modulation of the HBT radii with $\phi$ provides
five measurable quantities which over-determinates the four physical
quantities describing the source: its size $R$, deformation $\delta$,
opacity $(g_o-g_s)$ and duration of emission $\delta\tau$, at each
impact parameter. The azimuthal dependence of the HBT radii thus
offers an unique way to determine the opacity of the source as well as
the duration of emission separately.

Tracking these physical quantities with centrality will provide
detailed information about the source created in relativistic nuclear
collisions and may reveal a phase transition as sketched in Fig. (5). 

\acknowledgments 
Discussions with G. Baym, I. Bearden, J.J. G\aa rdh\o je,
U. Heinz, P. Huovinen, R. Mattiello, L. McLerran,
D. Rischke, V. Ruuskanen and S. Voloshin 
are gratefully acknowledged. Special thanks to A. Poskanzer for comments
and the improved NA49 data \cite{NA49v2}. 

\vspace*{20mm}
\appendix{ {\bf APPENDIX A: Evaluation of the Collision Integral} }\\

In order to calculate the elliptic flow, we need the particle distribution 
integrated over space for normalization in Eq. (\ref{v}).
With the free streaming distribution function of Eq. (\ref{fs}) we find
at any time $\tau$
\bea
 \frac{dN_1}{dyd^2p_\perp} &=& (2\pi)^{-3}\int 
   m_\perp \cosh(\eta-y) n(r,p,t) \tau d\eta d^2r_\perp    \nonumber\\
  &=&  (2\pi)^{-3} \tau_0 \int  dp_z \, f_0(p_\perp,p_z) \,. 
\label{dNdy}
\eea

The change of particle momenta due to collisions is the
change in the particle distribution function integrated over space and time
\bea
 \frac{d \Delta N_1}{dyd^2p_\perp} &=&  \frac{E_1}{(2\pi)^3}
 \int \left(\frac{\partial n}{\partial t}\right)_{coll} \,dt\,d^3r \,.
\eea
In the collisionless limit we can now insert the free streaming
distribution function (\ref{fs}) in the collision term in order to
calculate the first order correction to the distribution function. As
will be shown below, the system expands rapidly to low densities,
where collisions mainly contribute, and therefore the final state
factors $(1\pm n_i)$ from stimulated emission/Pauli blocking can be ignored.
For small angle scatterings the loss and gain terms in
$(n_3n_4-n_1n_2)$ nearly cancel leading to a suppression
factor at forward and backward angles $\sim(1-\cos^2\theta)$. 
Furthermore, little momentum is lost in small angle deflections leading
to little deformation in momentum space.
We therefore replace the cross section by an angular and energy
averaged {\it transport} cross section $\sigma_{tr}$
and keep only the loss term $\propto n_1n_2$
\bea
  \frac{d \Delta N_1}{dyd^2p_\perp} &=& -\frac{E_1}{(2\pi)^3} 
 \int d^2r_\perp \tau d\tau d\eta \int \frac{d^3p_2}{(2\pi)^3}
  \, v_{12} \sigma_{tr} \, n_1n_2 \,, \nonumber\\
 && \label{DN}
\eea
where we have changed variables to
the space-time rapidity and invariant time.

At late times we can utilize that the free streaming
distribution longitudinally can be approximated by
\bea
  f_0(p_\perp,p_z') \simeq \frac{\tau_0}{\tau} m_\perp^{-1}\delta(y-\eta) 
     \int dp_z f_0(p_\perp,p_z) \,. \label{f0}
\eea
where the latter integral can be eliminated by use of Eq. (\ref{dNdy}).
The resulting free streaming distribution function of Eq. (\ref{fs}) becomes
\bea
   n(x,p) &\simeq& S_\perp({\bf r}_\perp-{\bf v}_\perp\tau')
  \frac{(2\pi)^3}{\tau m_\perp} \delta(y-\eta) \frac{dN}{dyd^2p_\perp}
 \,. \label{nlate}
\eea

First we deal with the integration over transverse coordinates
\bea
 I_\perp &\equiv& \int d\rx d\ry \,
 S_\perp({\bf r}_\perp-{\bf v}_{1\perp}\tau') \,
 S_\perp({\bf r}_\perp-{\bf v}_{2\perp}\tau')  \nonumber\\
 &=&  \frac{1}{4\pi R_xR_y}
 \exp\left[ - \frac{\tau'^2}{4R_x^2}(v_{1x}-v_{2x})^2
 - \frac{\tau'^2}{4R_y^2}(v_{1y}-v_{2y})^2  \right] \,, \nonumber
\eea
with $\tau'=\tau-\tau_0$.
The transverse particle momentum or velocity with respect to the
reaction plane ($\rx$-axis) is ${\bf v}_{1\perp}=(v_{1x},v_{1y})=
v_{1\perp}(\cos\phi,\sin\phi)$, in terms of the 
azimuthal angle $\phi$.
Assuming deformations are small we obtain by expanding in $\delta$
\bea
 I_\perp &=&\frac{1}{4\pi R_xR_y}
 \exp\left[-\frac{\tau'^2 v_{12}^2}{4\bar{R}^2} \right] \nonumber\\
 &\times&
  \left(1-\frac{\tau'^2 v_{1\perp}^2}{4\bar{R}^2}  \delta\cos(2\phi)
 \,+\, {\cal O}(\delta^2) \right) \,.  \label{I}
\eea
where $v_{12}=|{\bf v}_{1\perp}-{\bf v}_{2\perp}|$ is the relative
velocity.  At this point we make the important observation that the
asymmetric term proportional to $\cos(2\phi)$ is weighted by a factor
$\tau'^2=(\tau-\tau_0)^2$, i.e., scatterings at early times contribute
little to azimuthal asymmetries in the momentum distribution whereas
they are important around $\tau\sim 2\bar{R}/v_{12}$. This justifies
Eqs. (\ref{f0}) and (\ref{nlate}) as well as the neglect of the Bose
and Fermi factors $(1\pm n_i)$ in the collision term as densities are
low at late times.
The physical reason for the ``late time'' dominance in elliptic flow is
that particles have to travel a distance $\sim R$ before it feels
the source deformation, i.e., whether it moves in the $\rx-$ direction,
where $R_x<R$, or in the $\ry-$ direction, where $R_y>R$.

Since $\tau_0\ll \bar{R}/v_\perp$, the time integral becomes
\bea
 \int_{\tau_0}^\infty\frac{d\tau}{\tau} \, I_\perp
 &=& - \frac{\delta}{8\pi R_xR_y} \frac{v_{1\perp}^2}{v_{12}^2}
 \cos(2\phi) +  constant   \,,  \label{It}
\eea
where the constant is independent of azimuthal angle $\phi$.
Finally, we carry out the $p_2$
and $d\eta$ integrals in Eq. (\ref{DN}) using (\ref{It}) with the result
\bea
 \frac{d \Delta N_1}{dyd^2p_\perp} &=& \langle v_{12}\sigma_{tr}\rangle
 \frac{dN_2}{dy}\frac{dN_1}{dyd^2p_\perp} \frac{\delta}{8\pi R_xR_y}
 \frac{v_{1\perp}^2}{\langle v_{12}^2\rangle}
 \cos(2\phi)  \nonumber\\
 &+&  constant   \,,  \label{const}
\eea
where the averaging over scatterer momenta is indicated by $\langle
.. \rangle$.  Originally we only included loss terms in the Boltzmann
equation and thereby replaced the cross section by the transport cross
section. The gain terms will not affect the asymmetric term but will
cancel the constant term in Eq. (\ref{const}) when the number of
particles is conserved. If there is net particle absorption or
production, $v_2$ should be multiplied by the ratio of initial and
final particle number.

We can now generalize to several kind of scatterers by replacing
$1$ by a given particle species $i$ as pions, protons, etc. and
replace the scatterer $2$ by a sum of scatterers $j$.
By dividing Eq. (\ref{const}) by $dN_1/dyd^2p_\perp$ we obtain 
by comparing to the definition, Eq. (\ref{v}), of elliptic flow 
\bea
 v_2^i = \frac{\delta}{16\pi R_xR_y}
  \sum_j \langle v_{ij}\sigma_{tr}^{ij} \rangle  \frac{dN_j}{dy} 
  \frac{v_{i\perp}^2}{{v_{i\perp}^2+\langle v_{j\perp}^2\rangle}}
     \,  \,. \label{v2ca}
\eea

\vspace{1cm}
\appendix{ {\bf APPENDIX B: Evaluation of equipotential flow in semi-central 
collisions} }\\

In this appendix we evaluate the elliptic flow for
an asymmetric source with 
equipotential flow which is parametrized spatially as a gaussian in 
transverse directions. We assume particles are thermally distributed
with local flow velocity ${\bf u}_\perp$ as in Eq. (\ref{u}).
By assumption the transverse flow is constant on the freeze-out
surface which is determined by
\bea 
  a^2 =  \frac{\rx^2}{R_x^2} + \frac{\ry^2}{R_y^2} \,. \label{ell}
\eea
i.e., $u_\perp$ is a function of $a$ but independent of the azimuthal angle
$\phi'=\tan^{-1}(\ry/\rx)$.
Also, the transverse flow velocity is perpendicular to the
elliptic surface
\bea
  {\bf u}_\perp &=& u_\perp  {\bf n}
 = u_\perp(a) \frac{\left( R_y^2\cos\phi',R_x^2\sin\phi' \right)}
  {\sqrt{R_y^4\cos^2\phi'+R_x^4\sin^2\phi'} }\,.
\eea 

In calculating the azimuthal dependence only the transverse directions 
are important as normalizations cancel in Eq. (\ref{v}). Thus
\bea
 \frac{dN}{dyd^2p_\perp} \propto \int a da  S_\perp(a) \int d\phi'
 \, \exp({\bf u}_\perp\cdot {\bf p}_\perp/T) \,.
\eea
Expanding for small deformations we obtain
\bea
 \frac{dN}{dyd^2p_\perp} &\propto& 
  \langle I_0(\frac{p_\perp u_\perp}{T}[1+\delta\cos(2\phi)]) \rangle  \,,
\eea
where the average  $\langle ..\rangle$ refers to radial average over $a$
of the transverse flow. $I_0$ is the  
Bessel function of imaginary argument which has the limits
\bea
  I_0(x) =  \left\{ \begin{array}{lrl}
     1+\frac{1}{4} x^2 & , & x\ll 1 \\
   \exp(x)/\sqrt{2\pi x}   & , & x\gg 1 \end{array} \right\} \,.\label{Bessel}
\eea           
Expanding for small deformations and
comparing to the definition of elliptic flow (\ref{v}) we finally obtain
\bea
  v_2 =  \left\{ \begin{array}{lrl}
    \frac{1}{4} \frac{p_\perp^2 \langle u_\perp^2\rangle}{T^2} \delta 
    & , & p_\perp u_\perp\ll T \\
    \frac{1}{2} (\frac{p_\perp \langle u_\perp\rangle}{T}-1) \delta 
      & , & p_\perp u_\perp\gg T  \end{array} \right\} \,.
\eea  

\vspace{1cm}
\appendix{ {\bf APPENDIX C: Apparent Temperatures and Transverse Flow} }\\

Transverse flow affects the measured $p_\perp$ and $m_\perp$
slopes. The {\it apparent} temperatures, obtained by fitting the
particle spectra by $dN/d^2p_\perp\propto \exp(-m_\perp/T_{app})$, are
larger than the intrinsic ones. It is difficult to determine
the intrinsic temperature and the transverse flow separately from
$p_\perp$ slopes of pions alone \cite{Schnederman}. 
Recent measurements \cite{NA44slopes} of apparent
temperatures for various massive particles, $\pi,K,p,d,^3He$, etc, may
allow us to estimate the transverse flow uniquely as will now be
described.

Assume a thermal source in two dimensions with intrinsic temperature $T$ 
at freeze-out and
(transverse) flow $u_\perp$ locally. i.e., $n\sim exp(p\cdot u/T)$. 
For small transverse deformations
the flow is almost azimuthally symmetric and its average value is equal
to the cylindrical symmetric value, $u_\perp(r)$. The distribution of
particles is thus
\bea
  n(r,p,t) \sim \exp[\frac{-\gamma m_\perp 
  +{\bf p}_\perp\cdot {\bf u}_\perp}{T}] \,. \label{nflow}
\eea
In three dimensions the source is further complicated by the thermal factor
$\exp(-m_\perp\cosh(y-\eta)/T)$. 
After integrating over longitudinal direction
or $\eta$, however, the result are very similar when $m_\perp\gg T$.

The dependence on the angle $\theta$ between 
${\bf p}_\perp$ and ${\bf u}_\perp$ in Eq. (\ref{nflow}) is crucial.
Integrating over transverse coordinates gives the distribution 
\bea
 \frac{dN}{d^2p_\perp} \propto \exp[-\gamma\frac{m_\perp}{T}] 
  \langle I_0(p_\perp u_\perp/T) \rangle  \, ,
\eea
where $\langle ...\rangle$ refers to radial average of the transverse flow
and $I_0$ the Bessel function (see Eq.(\ref{Bessel}).

The apparent temperature defined as the inverse $m_\perp$ slope becomes
\bea
  T_{app} &\equiv& [-\frac{d}{dm_\perp} \ln( \frac{dN}{d^2p_\perp})]^{-1}
   \nonumber\\
   &=& \left\{ \begin{array}{lrl}
    T + \frac{1}{2} m_\perp  \langle u_\perp^2\rangle
    & , & p_\perp u_\perp\ll T \,,\,m_\perp u_\perp^2\ll T\\
     T \sqrt{\gamma-u_\perp} 
      & , & p_\perp u_\perp\gg T \,,\,p_\perp\gg m \end{array}\right\} 
   \,. \nonumber \\
  &&  \label{Tap}
\eea
At small $p_\perp$ the result is the expected one when the kinetic energy 
of flow is added to the thermal energies in two dimensions. At large $p_\perp$
and flow one instead obtains the {\it blueshift} formula \cite{Vesa}.
Experimentally, the apparent temperature is determined by exponential
fits to $dN/d^2p_\perp$ in a certain region of $p_\perp$ and it may
therefore differ somewhat from (\ref{Tap}).

One should notice that in experimental fits to particle spectra the
apparent temperatures are parametrized as $T_{app} = T+m\beta^2$.
Consequently, this flow parameter differs from ours by a factor:
$\beta^2=\langle u_\perp^2\rangle/2$.


\end{multicols}

\clearpage

\vspace{-5cm}
\begin{figure}
\centerline{
\psfig{figure=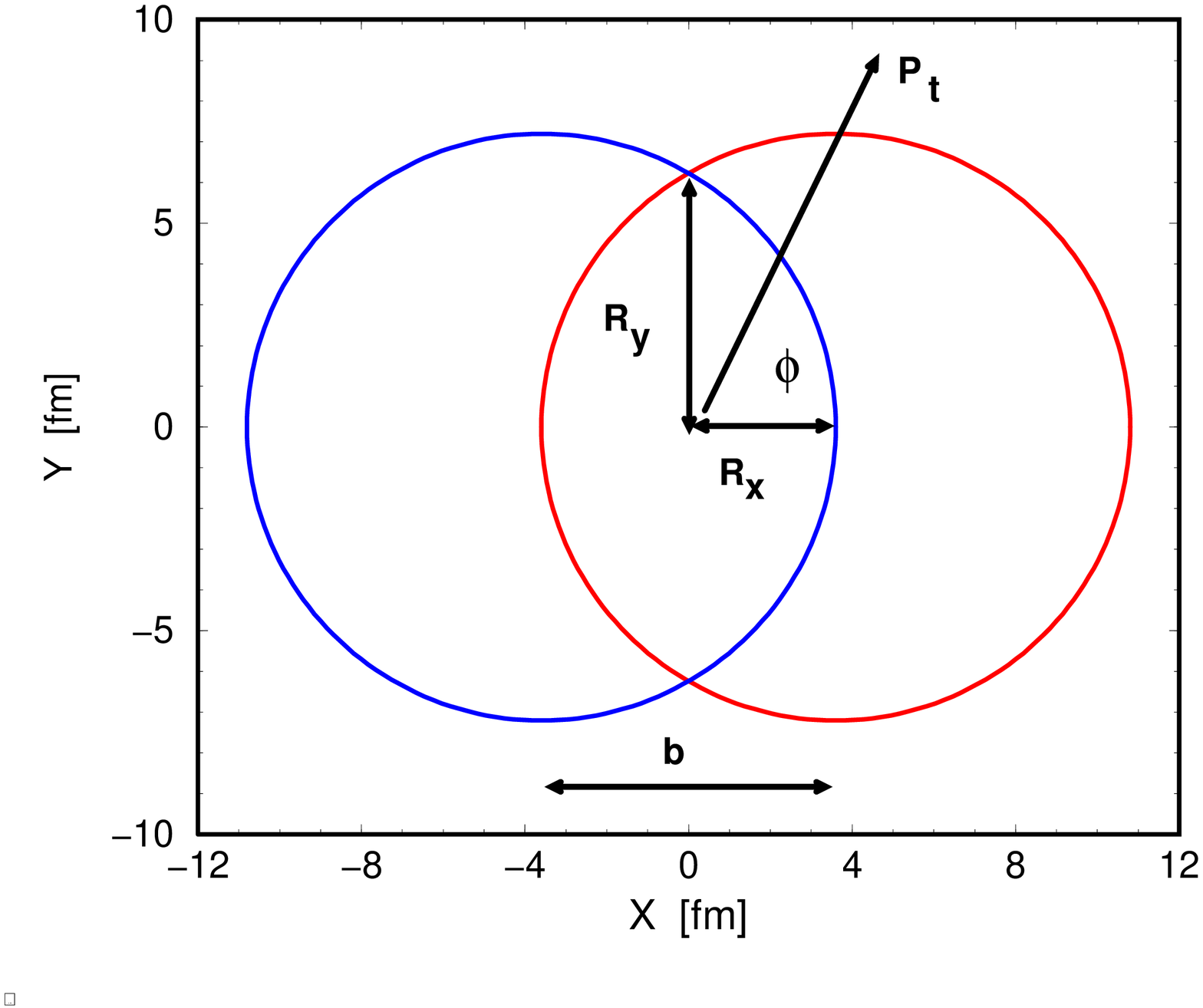,width=12cm,height=10cm,angle=-90}}
\vspace{1cm}
\caption{Reaction plane of semi-central $Pb+Pb$ collision for impact parameter
$b=R_{Pb}\simeq7$fm. The overlap
zone is deformed with $R_x\le R_y$. The reaction plane $(x,z)$ is
rotated by the angle $\phi$ with respect to the transverse particle momentum
$p_\perp$ which defines the outward direction in HBT analyses.
\label{geofig}  }
\end{figure}

\vspace{-2cm}
\begin{figure}
\centerline{\psfig{figure=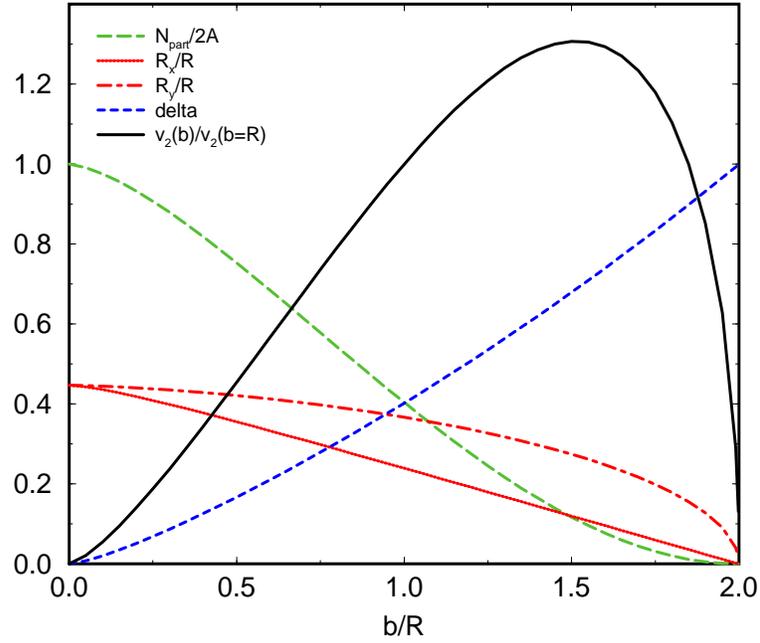,width=12cm,height=10cm,angle=-90}}
\vspace{1cm}
\caption{Transverse radii of nuclear overlap, 
deformation, number of participants and elliptic flow parameter (see
Eq.(\ref{v2c}) and text) versus impact parameter.
\label{bfig}  }
\end{figure}

\vspace{-5cm}
\begin{figure}
\centerline{\psfig{figure=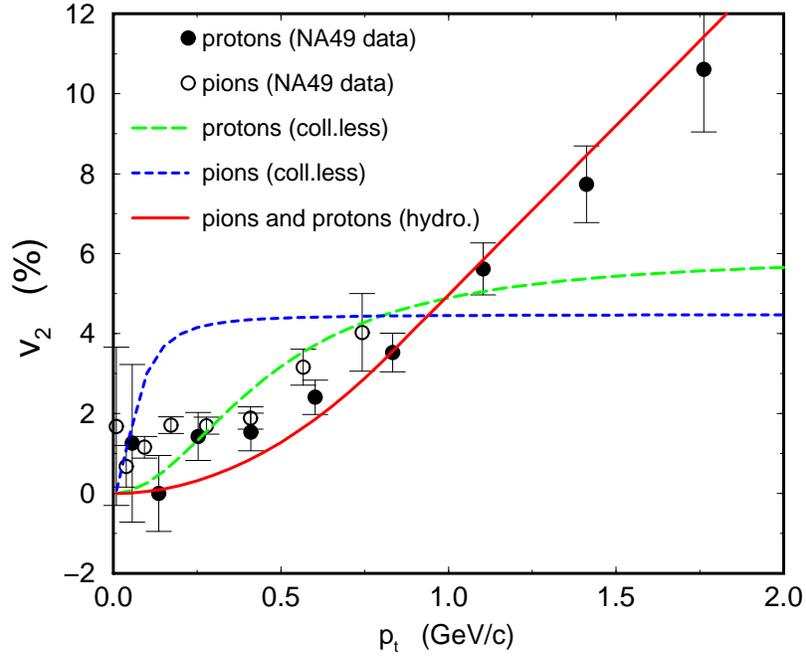,width=12cm,height=10cm,angle=-90}}
\vspace{0.5cm}
\caption{Elliptic flow $v_2$ for pions and protons
versus transverse momentum. Data from
NA49 \protect\cite{NA49v2}. Curves represent 
hydrodynamic limit of Eq.(\ref{v2h}) (full curve) and
collisionless limit of Eq.(\ref{v2c}) for
pions (dashed) and protons (dotted) (see text for details). 
\label{v2fig}  }
\end{figure}

\vspace{-2cm}
\begin{figure}
\centerline{\psfig{figure=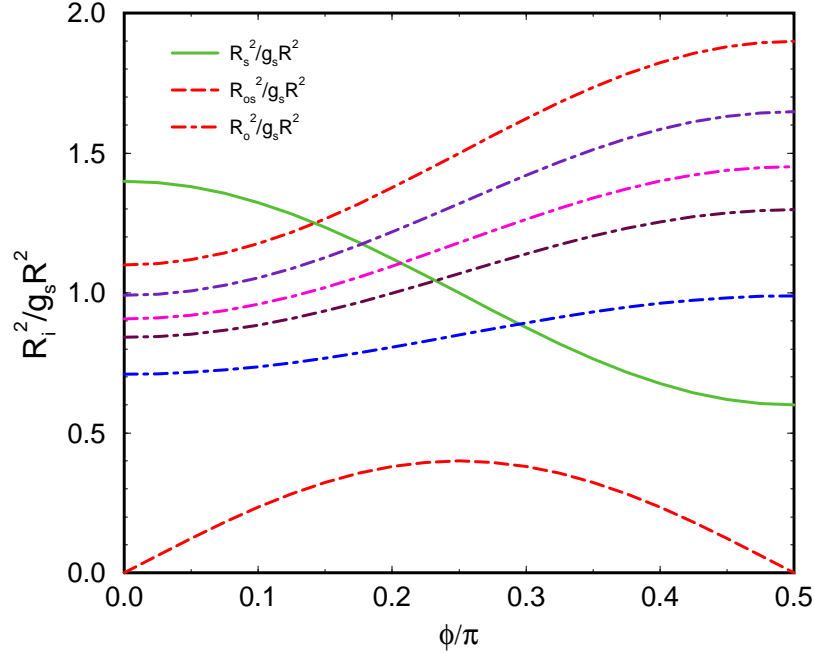,width=12cm,height=10cm,angle=-90}}
\vspace{0.5cm}
\caption{HBT radii vs. angle between reaction plane and transverse
particle momenta. The HBT radii are normalized 
to the angle averaged sideward HBT radius
squared, $g_sR^2$, for deformed $\delta=0.4$ source with duration
of emission $\beta_o^2\delta\tau^2/g_sR^2=0.5$ and with various opacities. 
The sideward HBT radius (full curve) is then the
same for both transparent (Eq. (\ref{Rst})) and opaque (Eq. (\ref{Rso}))
sources and likewise for the out-side HBT radii (Eqs. (\ref{Rost}) and 
(\ref{Roso}), dashed curve). 
The outward HBT radii are shown with  chain-dashed curves
for a gaussian source (Eq. (\ref{Roo}) and (\ref{Rot}) with various
opacities (from below and up): $\lambda_{mfp}/R=0.1,0.5,1.0,2.0,\infty$.
\label{R2fig}  }
\end{figure}

\vspace{-1cm}
\begin{figure}
\centerline{\psfig{figure=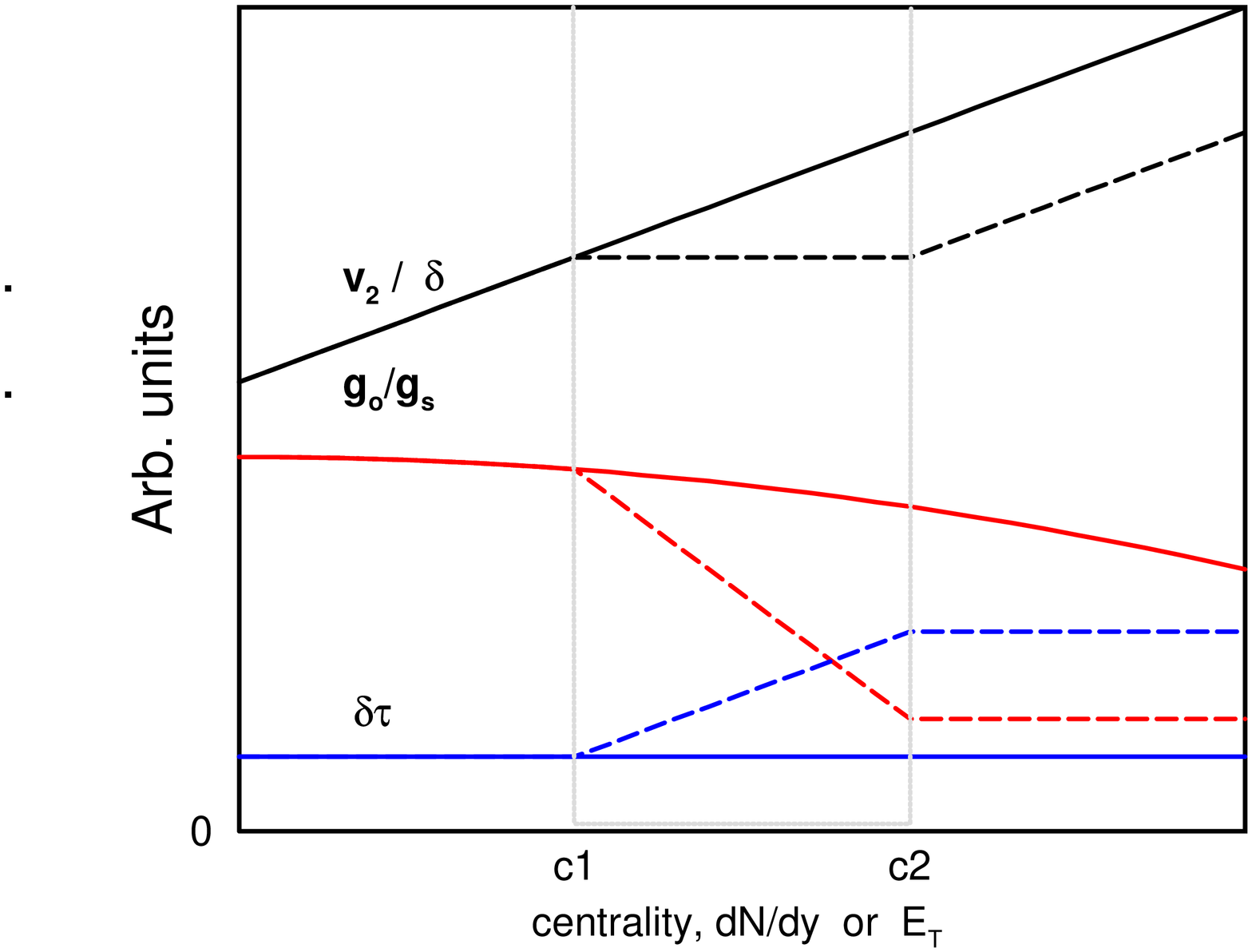,width=12cm,height=10cm,angle=-90}}
\vspace{1cm} 
\caption{Skematic behavior of the centrality, $dN/dy$ or $E_T$ 
dependence with and without a phase transition. From top to bottom
the curves show the ratio of elliptic flow to deformation $v_2/\delta$,
opacity $g_o/g_s$, and duration of emission $\delta\tau$. The expected
behavior without (full curves) and with (dashed curves) a first order
phase transition occurring in a region of semi-centrality from $c1$ to $c2$
is described in the text. \label{fig5}  }
\end{figure}


\begin{references} 

\bibitem{Bevalac} W. Reisdorf and H.G. Ritter, {\it Ann. Rev. Nucl. Part. Sci.}
                  {\bf 47} (1997) 663.
\bibitem{AGS} E877 coll., J. Barrette et al.,  {\it Phys. Rev.}
                  {\bf C56} (1997) 3254; E895 coll., H. Liu et al., 
{\bf A638} (1998) 451c.
\bibitem{NA49v2} H. Appelsh\"{a}user et al. (NA49), 
    {\em Phys. Rev. Lett.} {\bf 80}, 4136 (1998). 
    A. Poskanzer (NA49), {\it Nucl. Phys.} {\bf A638} (1998) 463c.
 Improved data from na49info.cern.ch home page has been employed.

\bibitem{HBT} R. Hanbury--Brown and R.Q. Twiss, {\em Phil. Mag.} {\bf
		45} (1954) 633.
\bibitem{Voloshin} S.A. Voloshin, {\it Phys. Rev.} {\bf C55} (1997) R1630;
A.M. Poskanzer and S.A. Voloshin, {\it Phys. Rev.} {\bf C58} (1998) 1671.
\bibitem{Wiedemann} U. Wiedemann, {\it Phys. Rev.} {\bf C57}, 266 (1998). 
\bibitem{Ollitrault} J.Y. Ollitrault, {\it Phys. Rev.} {\bf D46} (1992) 229;
                    {\it ibid.} {\bf D48} (1993) 1132; and Proc. of QM'97.
\bibitem{RQMD} H. Sorge, {\it nucl-th/9610026}.
\bibitem{Xu} H. Liu, S. Panitkin, and N. Xu, {\it nucl-th/980721}.
\bibitem{NA49} T. Alber et al., {\em Z. Phys.}  {\bf C 66} (1995) 77;
    K. Kadija et al., {\em Nucl. Phys.} {A 610} (1996) 248c; 
    R. Ganz et al., {\it nucl-ex/9808006}. 
\bibitem{Barz} H.W. Barz, J. Bondorf, J.J. G\aa rdh\o je, and H. Heiselberg, 
         {\it Phys. Rev.} {\bf C56} (1997) 1553; 
         {\it ibid} {\bf C57} (1998) 2536. 
\bibitem{Nordheim} L.W. Nordheim, {\it Proc. Roy. Soc.} (London) V,
       {\bf A119} 689 (1928).
\bibitem{Baym} G. Baym, {\it Phys. Lett.} {\bf 138B}, 18 (1984).
  H. Heiselberg and X.-N. Wang, {\it Phys. Rev.} {\bf C53}, 1892 (1996).
\bibitem{Bhydro} G. Baym, B. Friman, J.-P. Blaizot and M. Soyeur,
   {\it Nucl. Phys.} {\bf A407}, 541 (1983).
\bibitem{Kataja} M. Kataja, P.V. Ruuskanen, L.D. McLerran, H.von Gersdorff,
                    {\it Phys. Rev.} {\bf D34} (1986) 2755.
\bibitem{NA44slopes} I.G. Bearden et al. (NA44 collaboration), 
		{\em Phys. Rev. Lett.} {\bf 78} (1997) 2080. 
\bibitem{Bertsch} H.W. Barz, G. Bertsch, P. Danielewicz and H. Schulz,
  {\it Phys. Lett.} {\bf B 275}, 19 (1992).
\bibitem{HBD} H. Heiselberg,  nucl-th/9809077, 
to appear in {\it Phys. Rev. Lett.} (1999).
\bibitem{NA44} A. Franz, {\em Nucl. Phys.} {\bf A 610} (1996) 240c.
               I.G. Bearden et al., {\it Phys. Rev.} {\bf C58} (1998) 1656. 
\bibitem{GKW} M. Gyulassy, S.K. Kaufmann and L.W. Wilson,
              {\it Phys. Rev.} {\bf C 20} (1979) 2267.
\bibitem{Pratt} S. Pratt, {\em Phys. Rev. Lett.} {\bf 53} (1984) 1219.
\bibitem{Csorgo} T. Cs\"org\H o and B. L\"orstad, {\em Nucl. Phys.} 
		{\bf A590}, 465c (1995); {\em Phys. Rev.} {\bf C54}
 		(1996) 1390.
\bibitem{Heinz} S. Chapman, J.R. Nix, and U. Heinz, {\em Phys. Rev.} 
		{\bf C52}, 2694 (1995).
\bibitem{HH} H. Heiselberg, {\em Phys. Lett.} {\bf B379} (1996) 27.
   U.A. Wiedemann and U. Heinz, {\it Phys. Rev.} {\bf C56} (1997) 3265. 
\bibitem{Opaque} H. Heiselberg and A.P. Vischer, {\it Eur. Phys. J.} {\bf C1},
 593 (1998); {\it Phys. Lett.} {\bf B421} (1998) 18; 
\bibitem{Sinyukov} Yu. M. Sinyukov, V.A. Averchenkov, B. L\"orstad,
                   {\em Z. Phys.} {\bf C49} (1991) 417.
\bibitem{Tomasik} B. Tomasik and U. Heinz, nucl-th/9805016.
\bibitem{CF} F. Cooper and G. Frye, {\em Phys. Rev.} {\bf D10} (1974) 186.
\bibitem{Bugaev} K.A. Bugaev, {\it Nucl. Phys.} {\bf A606} (1996) 559.
\bibitem{Lattice} C. Bernard et al., {\it Phys.Rev.} {\bf D55} (1997) 6861.
\bibitem{vanHove} L. Van Hove, {it Phys. Lett.} {\bf 118B} (1982) 138.
\bibitem{CalCurve} J. Pochodzalla et al. (ALADDIN), 
{\it Phys. Rev. Lett.} {\bf 75}, 1040 (1995). 
\bibitem{Hauger} J.A. Hauger et al., {\it Phys. Rev. Lett.} {\bf 77}
(1996) 235.
\bibitem{Bondorf} J.P.Bondorf, R. Donangelo, I.N. Mishustin and H. Schulz, 
{\it Nucl. Phys.} {\bf A444} (1985) 460.
\bibitem{Rischke} S. Bernard, D.H. Rischke, J.A. Maruhn, W. Greiner
                  {\it Nucl. Phys.} {\bf A625}, 473 (1997). 
\bibitem{HJ} H. Heiselberg and A.D. Jackson, {\it nucl-th}/9809013.
\bibitem{J} H. Heiselberg and R. Mattiello,  {\it nucl-th}/9901004.
\bibitem{Schnederman} E. Schnedermann and U. Heinz,
                      {\it Phys. Rev. Lett.} {\bf 69} (1992) 2908. 
\bibitem{Vesa} P.V. Ruuskanen, {\it Z. Phys.} {\bf C38} (1988) 219.

\end{references}
\end{document}